# Multi-Mode Lens for Momentum Microscopy and XPEEM: Theory

Olena Tkach and Gerd Schönhense

*Johannes Gutenberg-Universität, Institut für Physik, 55128 Mainz, Germany*

## Abstract

The strong electric field between the sample and the extractor is the core of cathode lenses and a pivotal determinant of high resolution. Nevertheless, fields in the range of 3-8 kV/mm can be a source of complications. Local field enhancement at sharp edges or microscopic protrusions of cleaved samples may result in field emission or flashovers. Moreover, slow background electrons are drawn into the microscope column, where they contribute to space charge effects. A novel front lens configuration, optimized through ray-tracing simulations, significantly reduces the field at the sample and allows even for zero field or retarding field, which serves to suppress space charge effects. One or several annular electrodes, situated in a concentric position relative to the extractor, serve to form an additional lens within the gap between the sample and the extractor. The refractory power of this lens, and consequently the field at the sample surface, can be modified by adjusting the potentials of the annular electrodes. The imaging properties and aberrations of this gap lens have been investigated with regard to momentum imaging and XPEEM. The study encompasses the energy range from the few-eV level for laser-ARPES to 6 keV, for hard X-ray ARPES. The additional converging lens situated in close proximity to the sample exhibits a reduced field curvature of the *k*-image in the backfocal plane. This allows for the acquisition of larger fields of view in both momentum and real-space imaging.

## 1. Introduction

The performance of cathode lens microscopes (PEEMs, LEEMs and MMs) is contingent upon the strength of the accelerating field between the sample and the extractor. The simplified treatment of Brüche and Recknagel [1,2] has subsequently been refined through extensive theoretical work by Bauer and Rempfer et al. [3,4]. Ultimately, Tromp et al. derived the aberration coefficients of cathode lenses up to the fifth order [5,6]. Typical cathode lenses of PEEMs and MMs are designed for extractor fields in the range of 3-8 kV/mm. Such fields can prove problematic for cleaved samples, which often exhibit sharp corners or protruding lamellae, with the potential for field emission or even flashovers.

A second issue can emerge from the space charge interaction. The extractor field attracts slow electrons into the lens column, where they exert Coulomb forces on the photoelectrons. The number of slow secondary electrons, resulting from cascade scattering processes, can potentially exceed the true photoelectrons by several orders of magnitude. This is particularly problematic in experiments utilizing pulsed X-ray beams from synchrotrons or free-electron lasers. The space-charge effect has been observed to impose a resolution limit in X-ray excited PEEM (XPEEM) [7-9] and to induce a deformation (Lorentzian shaped) of the isoenergetic planes in momentum microscopy [10]. In time-resolved photoemission experiments employing the pump-probe technique, another species of slow electrons can be released by the pump pulse via nPPE transitions. The long-range Coulomb interaction between this species and the photoelectrons results in artefact signals that may obscure the true effects [11].

The strong extractor field has thus been identified as a potential source of complications in several instances. The prevailing notion that high extractor fields are essential for the optimal cathode lens functioning is often met with scepticism, given the practical constraints that impede their application. This prompted our investigation into alternative configurations of the front lens, with the objective of circumventing the strong extractor field.

In the novel lens configuration, the homogeneous accelerating field is replaced with a *gap lens*, which is an additional lens situated in the space between the sample and the first electrodes. The potential distribution is formed by one or two annular electrodes that are concentric and coplanar with the extractor. The ring electrodes do not impede the space between the sample and the extractor, yet they are capable of generating a variety of lens fields within the gap. The application of different voltages enables the field at the sample surface to be tuned from a strongly accelerating to a strongly retarding configuration. Each of which has a different effect on the imaging behavior.

The diverse operational modes of the gap-lens geometry are investigated via ray-tracing simulations. Notwithstanding the diminished field at the sample (thereby obviating the risk of field emission), the field curvature of the *k*-image in the backfocal plane is markedly reduced in comparison to the extractor mode. This enables significantly larger *k*-fields of view with diameters up to 18 $Å^{-1}$, which is advantageous for photoelectron diffraction, and up to > 4 mm in PEEM mode, which is a crucial aspect for the forthcoming full-field imaging PAXRIXS technique [12]. Tuning the field at the sample to zero is advantageous for specimens exhibiting corrugations or electrodes on the surface. The application of retarding fields serves to repel slow electrons, thereby eliminating the major part of the vacuum space-charge interaction in pump-probe experiments (time-resolved ARPES). The ray-tracing results are presented and discussed here, with the initial experimental results shown in Ref. [13].

## 2. The multi-mode front lens

### *2.1 Motivation for the development of the multi-mode lens*

The objectives of PEEMs and MMs are distinct and therefore the two instruments are optimized for different purposes. The objective of PEEMs is to achieve high-resolution imaging in real space, whereas MMs are designed for reciprocal-space imaging, whereby $k_x$-$k_y$ patterns are recorded. In accordance with the specific application, the criteria for evaluating the quality of MM images may vary. In the case of recording valence band patterns, the quantity of interest is the *k*-resolution. Benchmark values in the range of 0.004 Å$^{-1}$, obtained with a double-hemispherical MM [14] are comparable to those obtained by ARPES using hemispherical analysers [15]. In time-of-flight (ToF) MM at higher energies in the soft or hard X-ray range [16,17] additional criteria become pertinent. In the context of X-ray photoelectron diffraction (XPD), the size of the momentum field of view is of paramount importance, as it determines the information content of the diffraction patterns. The recording of large *k*-fields, with diameters exceeding 12 Å$^{-1}$, has been demonstrated using hard X-ray excitation [18,19]. In the case of such large *k*-fields, the *spherical aberration* of the lens system represents a significant limiting factor.

In regard to *chromatic aberration*, the circumstances are contingent upon the specific instrument in question. In the case of MMs based on hemispherical analyzers, recording 2D momentum patterns ($k_x$,$k_y$), the bandwidth of the electrons detected is relatively narrow. In laboratory experiments utilizing a He lamp, the ultimate energy resolutions achieved by different instruments ranged from 4 to 13 meV FWHM [20-22]. In synchrotron experiments the resolution is typically constrained by the bandwidth of the photon beam, with typical values of 30-50 meV in the soft X-ray range and an order of magnitude larger for the hard X-ray range. In these conditions, the chromatic aberration is not a significant issue. ToF-based MMs [16,17] record a specific energy band simultaneously, thereby rendering chromatic aberration a significant factor. However, there are two potential applications: In full-field photoelectron diffraction on selected core levels the energy band of interest is narrow, with typical spectral ranges used for XPD being 500 meV. Then, the impact of chromatic aberration will remain insignificant. The situation is distinct for ToF-MM-based ARPES, wherein energy bands of several eV width are recorded as 3D data array ($k_x$,$k_y$,$E_{kin}$). In this instance, chromatic aberration imposes a limitation on the width of the usable energy interval. Two distinct effects are observed as signature of chromatic aberration. The *transversal chromatic aberration* (also termed *chromatic aberration of magnification*) results in a variation of the image magnification that can be readily rectified through numerical correction [23]. However, the *longitudinal chromatic aberration* results in an image blurring with increasing deviation from the selected center energy, which cannot be corrected. We notice that the aberrations of the objective lens can be compensated for by a mirror corrector, which requires an arrangement with two magnetic beam separators [24].

The development of the multimode lens was driven by the aforementioned issues caused by the high field strength (voltage breakthroughs) and by the space-charge effects (energy shifts and broadening, image blur) of cathode lens microscopes. The fundamental concept underlying the design of the new front lens was the ability to vary the field at the sample surface over a broad range, from accelerating to retarding, through the modification of lens potentials. The primary objective was to achieve optimal *k*-imaging quality with an accelerating field of reduced strength, a factor of 3-4 lower than typical fields of classical cathode lenses. In instances where the sample exhibits critical characteristics, such as sharp corners or protrusions, this mode should be employed in lieu of the extractor mode. The second objective was to develop a retarding mode that would eliminate the principal contribution to the space charge effect. A systematic ray-tracing approach was employed to investigate the behaviour of all modes across a range of lens geometries and kinetic energies. All simulations were conducted using the SIMION code [25].

Figure 1 depicts the front part of the novel lens, which is equipped with two annular electrodes designated as R1 and R2. The action of these electrodes is illustrated by five characteristic lens field distributions. The first examples of such electrodes were proposed in Ref. [26]. By modifying the potential of R1 and R2, additional degrees of freedom for field shaping can be introduced. The complete front lens comprises the sample (sa), the extractor electrode (Ex), one or two ring electrodes (R1 and R2) and the focusing lens situated further downstream. In the simulations, the gap sizes assumed were $\ell$ = 6 and 8 mm.

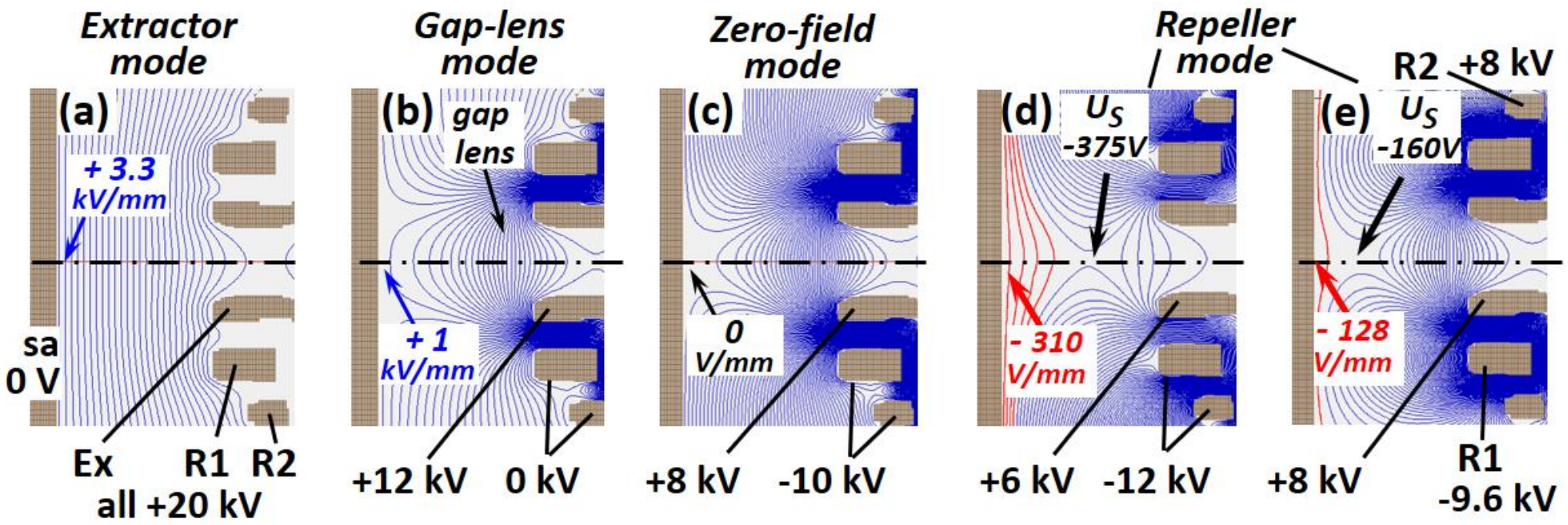

**Fig. 1.** Operating schemes for the various modes of the new front lens, with the contours of equipotential surfaces. (a) Classic *extractor mode* with the extractor Ex and ring electrodes R1 and R2 at 20 kV, resulting in a homogeneous field of $F$=+3.3 kV/mm. (b) *Gap-lens mode* with $U_{Ex}$= 12 kV and $U_{R1}$= $U_{R2}$ =0, resulting in the formation of an additional lens in front of Ex, which reduces the field to $F$= +1 kV/mm at the sample. (c) *Zero-field mode* ($F$= 0), achieved with $U_{Ex}$ = 8 kV and negative ring electrodes ($U_{R1}$=$U_{R2}$= -10 kV). (d) *Repeller mode*, here with $F$= -310 V/mm, attained with $U_{Ex}$= 6 kV and $U_{R1}$= $U_{R2}$= -12 kV. The saddle point at $U_S$ = -375 V defines the low-energy cut-off. (e) *Repeller mode*, modified by setting $U_{Ex}$= $U_{R2}$= +8 kV and $U_{R1}$= -9.6 kV. In (d,e) the retarding field is indicated by red potential contours.

When the extractor and both ring electrodes at set to high positive potential with respect to the sample (here $U_{Ex}$ =$U_{R1}$=$U_{R2}$= 20 kV), the configuration exhibits the characteristics of a classical cathode lens in *extractor mode*, Fig. 1(a). The homogeneous field is $F = U_{Ex}/\ell$, in the present case $F$ = +3.3 kV/mm for $\ell$ = 6 mm. By reducing the potentials of R1 and R2, an additional converging lens is introduced into the gap between the sample and the extractor. Figure 1(b) illustrates the equipotential contours for $U_{Ex}$= 12 kV and $U_{R1}$ = $U_{R2}$ = 0. In this *gap-lens mode* the accelerating field at the sample is significantly reduced, in the present case to +1 kV/mm. By setting R1 and R2 to negative voltages, it is possible to compensate the field at the sample surface. This establishes the *zero-field mode,* as illustrated in Fig. 1(c), for $U_{Ex}$ = 8 kV and $U_{R1}$ = $U_{R2}$ = -10 kV.

A further increase in the negative potentials at the ring electrodes results in the *repeller mode*, as illustrated in Fig. 1(d), with $U_{R1}$ = $U_{R2}$ = -12 kV. The retarding field is indicated by red equipotential contours, with a field strength at the sample of -310 V/mm. The saddle point at a potential of $U_S$ = -375 V functions as a high-pass filter, allowing only electrons with kinetic energies >375 eV to pass. The repulsion of electrons with lower energy results in a notable decrease in the Coulomb interaction within the beam. A further degree of freedom for field shaping is introduced by setting R1 and R2 to different potentials. It is possible to modify the repeller field in order to create different shapes, and the position of the saddle point can also be adjusted. In the illustrative example depicted in Fig. 1(e), wherein $U_{R1}$ is set to -9.6 kV and $U_{R2}$ to +8 kV, the retarding field is diminished, the saddle point is shifted closer to the sample and the refractive power of the gap lens is increased.

### *2.2 Gap-lens mode*

A principal objective of the present study was to examine the performance of the *gap-lens mode* (Fig. 1(b)), which serves to reduce the field at the sample surface. This section aims to elucidate the question of how the imaging quality of the gap-lens mode compares with that of the classical extractor mode, particularly when $E_{kin}$ increases. The simulations, together with the initial experimental validations, were conducted for a configuration comprising a single ring electrode (R).

Figure 2 provides a response to this question. Ray-tracing simulations were conducted for three distinct energies: $E_{kin}$= 150 eV, 900 eV and 6 keV. The former is an energy that can be achieved by upcoming high-harmonic-generation (HHG) based sources, while the latter two are typical soft and hard X-ray energies, respectively. The results for the *extractor mode* are pesented in Figs. 2(a,e and h). The application of 20 kV to the extractor and ring electrode results in the generation of a homogeneous field of 3.3 kV/mm in a 6 mm gap (Fig. 2(a,e)) and 2.5 kV/mm in an 8 mm gap (Fig. 2(h)).

A simulation of the full solid angle of $2\pi$ (0-90°) was conducted for 150 eV, resulting in a diameter of 12 $Å^{-1}$. As a consequence of the spherical aberration, the *k*-image is situated on a curved focal surface delineated by the dashed blue curve in (a). As the value of $k_{\parallel}$ increases, the rays become increasingly over-focused. In the backfocal plane (BFP), marked by the dotted red line, this phenomenon gives rise to an increasing degree of image blur with increasing distance from the centre. The detector is located in a conjugate plane of the BFP at a subsequent point along the optical path, where the *k*-image exhibits the same field curvature. The portion of the image that is suitable is reduced to approximately 50°, corresponding to a *k*-field of diameter 9.4 $Å^{-1}$. For extractor voltages lower than 20 kV, the curvature in (a) is even more pronounced, resulting in a further reduction in the usable angular range.

Figures 2(b) and (c) demonstrate the trajectories of the *gap-lens mode* for identical input parameters ($E_{kin}$ = 150 eV, solid angle $2\pi$). By setting the ring electrode to $U_R$ = -1 kV with $U_{Ex}$ still at 20 kV, the homogeneous field is replaced by an accelerating lens in the gap between the sample and the extractor, as indicated by the dotted ellipse in Fig. 2(c). The field at the sample surface is reduced from 3.3 kV/mm (a,d) to 1.6 kV/mm. As the distance from the sample increases, the field strength rises to a maximum of 3.6 kV/mm at the centre of the gap lens. The field curvature is markedly diminished, as evidenced by the flattening of the blue curve in Fig. 4(b). The full angular range of 0° to 90° is now accessible, although there is a slight reduction in *k*-resolution near the image rim. An increase in the negative voltage applied at R serves to further reduce the field curvature. Fig. 2(d) presents the same simulation for the complete ToF MM optics. The entire *k*-field with a diameter of 12.2 $Å^{-1}$ is focused on the detector (final *k*-image), with minimal loss of image quality.

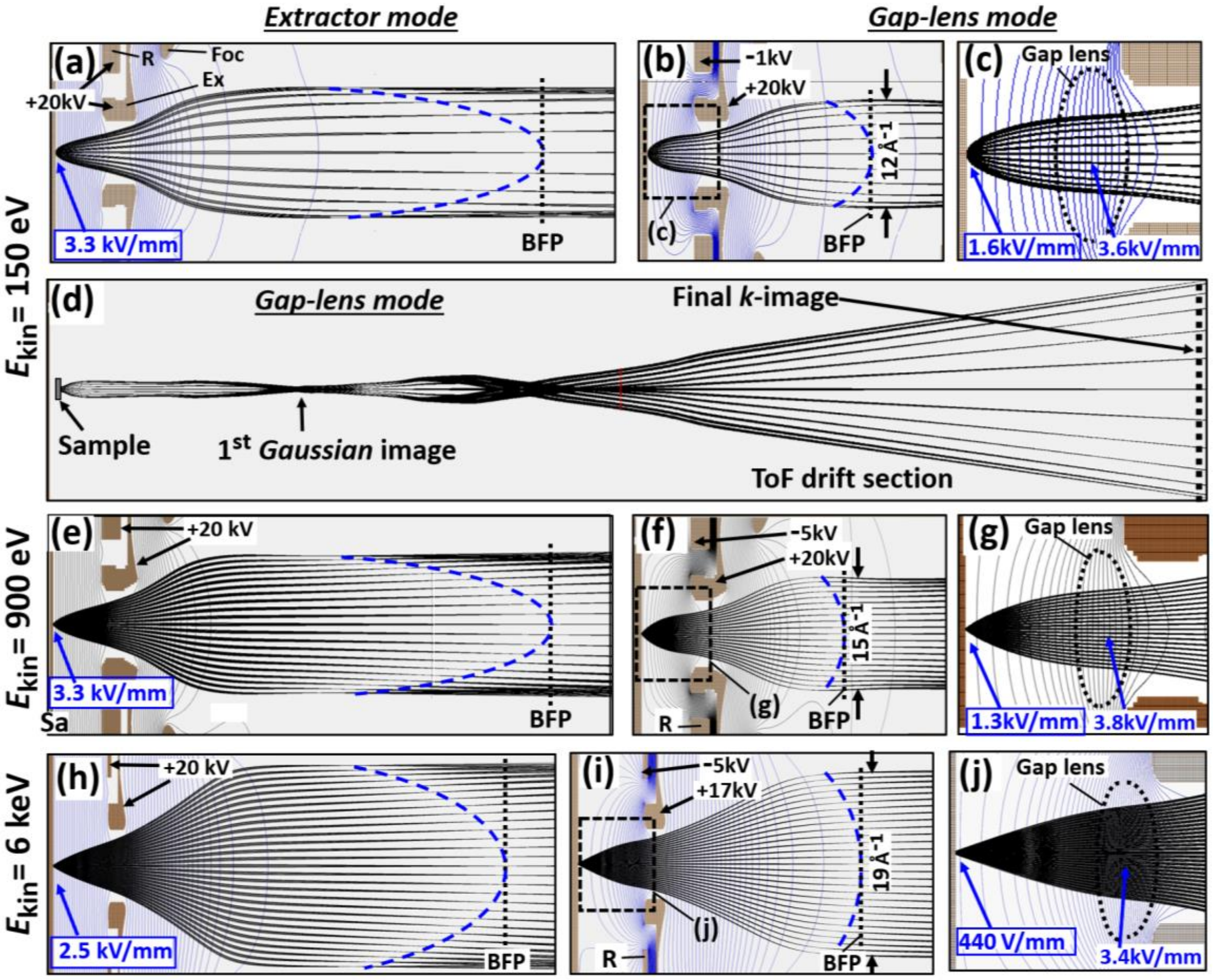

**Fig. 2.** Comparison of *extractor* and *gap-lens modes* for $E_{kin}$ = 150 eV, 900 eV and 6 kV; source spot diameter 100 μm. (a-c) Simulation for 150 eV and $\alpha$ = 0°-90°, corresponding to a *k*-field of 12.2 Å$^{-1}$ in diameter. (a) *Extractor mode* with $U_{Ex} = U_R$ = 20 kV. The backfocal plane (BFP) defined by the paraxial *k*-image is marked by the dotted line. The strong *k*-image curvature (dashed blue curve) due to the spherical aberration reduces the usable *k*-field. (b,c) *Gap-lens mode* with $U_R$ = -1 kV; detail (c) is marked as rectangle in (b). The field curvature is weaker than for the *extractor mode.* (d) Same for a complete momentum microscope optic; the full 2π solid angle at $E_{kin}$ = 150 eV is imaged on the planar detector. (e-g) and (h-j) analogous simulations for $E_{kin}$ = 900 eV ($\alpha$ = 0°-30°, 15 Å$^{-1}$ diameter) and 6 keV ($\alpha$ = 0°-14°, 18.7 Å$^{-1}$ diameter). $U_{Ex}$ = +20 kV in (e-h) and +17 kV in (i,j), $U_R$ = -5 kV in (f,g,i,j).

The results for $E_{kin}$ = 900 eV and $\alpha$ = 0°-30° are qualitatively similar. As a consequence of the larger *k*-field diameter of 15 Å$^{-1}$, the field curvature in the *extractor mode* results in a further blurring of the outer part of the image, as illustrated in Fig. 2(e). The usable angular range is thus reduced to 20°, corresponding to a *k*-field diameter of 10 Å$^{-1}$. The trajectories for the *gap-lens mode* with R set to -5 kV are illustrated in Figs. 2(f) and (g). In this instance, the complete field of 15 Å$^{-1}$ is imaged with an excellent level of quality on the planar detector. In accordance with these settings the field at the sample is 1.3 kV/mm and rises to 3.8 kV/mm in the center of the gap lens (Fig. 2(g)).

For $E_{kin}$ = 6 keV the gap size was increased to 8 mm. The simulation was performed for $\alpha$ = 0°-14°, which corresponds to a *k*-field diameter of 19 Å$^{-1}$. As in the preceding cases, the field curvature in the extractor mode results in a blurring of the the outer part of the image in the BFP (Fig. 2(h)). The angular range that can be utilized is approximately 9°, which corresponds to a *k*-field diameter of 12 Å$^{-1}$. The trajectories for the *gap-lens mode* with $U_{Ex}$ = 17 kV and $U_R$ = -5 kV are depicted in Figs. 2(i) and (j). The complete *k*-field of 19 Å$^{-1}$ is imaged with high quality on the planar detector. In this configuration, the field at the sample is as low as 440 V/mm and can be reduced to zero for larger negative potentials at R, as will be discussed in the following section. These settings are optimal for operando samples with contacts or actuators on the surface. In particular, the gap-lens mode is advantageous for hard X-ray photoelectron diffraction studies, as it provides access to *k*-field diameters of up to 20 $Å^{-1}$. However, imaging such large $k_{\parallel}$-values necessitates the incorporation of subsequent electron lenses with sufficiently large diameters, maintaining reasonable filling factors.

It can be concluded that the additional lens positioned between the extractor and the sample has the effect of reducing the field strength at the sample surface when compared to similar imaging conditions where the extractor mode is used. Moreover, the momentum-image quality is improved across the entire energy range from the XUV to the soft and hard X-ray regimes. Even for the full solid angle of $2\pi$ at $E_{kin}$ = 150 eV, the field curvature remains within acceptable limits, as illustrated in Figs. 2(b) and (d). The combination of a reduced field at the sample and improved image quality renders the gap-lens mode a promising option for many applications. This mode was employed in experiments in the soft X-ray range at beamline P04 at PETRA III (DESY, Hamburg, Germany) [27,28] and is frequently utilized at the HHG-based fs XUV source at CELIA, Bordeaux (first results in Ref. [13]).

### *2.3 Repeller and zero-field mode*

By setting the annular electrode R to a larger negative voltage and reducing $U_{Ex}$, it is possible to tune the field at the sample surface to zero or even to negative values. Such modes are designated as *zero-field* and *repeller modes*. The gap-lens mode (see Section 2.2) represents a flexible alternative to the conventional extractor mode. In contrast, the repeller mode is primarily designed for a specific application, namely the reduction of space charge interaction. Consequently, this mode is especially appealing for time-resolved photoelectron momentum microscopy utilizing pump-probe techniques. In these experiments, a substantial number of slow electrons can be emitted by the pump pulse through nPPE processes or plasmonic emission. Time-resolved ToF-MM at free-electron lasers and HHG-based laboratory sources represents an emerging technique with a significant gradient. Refs. [29-31] provide insight into several aspects of this technique.

Figure 3 illustrates the *repeller mode* for four kinetic energies between 36 eV and 3.6 keV. In the energy range between 36 eV and 3.6 keV the extractor potential is reduced from $U_{Ex}$ = +9.5 kV to +2.07 kV, resulting in repeller fields at the sample surface between *F* = -11 V/mm and -1.23 kV/mm, respectively. The retarding field determines the distance at which electrons with a given energy are redirected towards the sample surface. The saddle point functions as a high-pass filter, with the potential $U_S$ defining the cut-off energy for the electrons escaping from the sample. In the sequence from $E_{kin}$ = 36 eV to 3.6 keV, the cutoff energy increases from 3 eV to 3.15 keV, as evidenced by the values presented in the panels of Fig. 3. In repeller mode, the refractive power of the front lens is significantly increased; therefore the first *k*-image is located in the gap region and the first Gaussian image appears in the region of the extractor bore.

Fig. 3(a) depicts the front-lens region, while Fig. 3(b) illustrates the complete lens optics of a ToF-MM for an electron energy of 36 eV and an angular range of 0°-30° (*k*-field diameter 3 $Å^{-1}$). The simulation for the full ToF MM optics demonstrates that the minor aberration is sustained until the final *k*-image on the detector, i.e. the aberrations of the subsequent lenses are not significant. The application of $U_R$ = -12 kV and $U_{Ex}$ = +9.5 kV results in a field strength of $F$ = –11 V/mm at the sample surface and $U_S$ = -3 V. This configuration effectively removes all slow electrons with kinetic energies up to 3 eV from the beam. Fig. 3(c) depicts the particulars of the front lens for $E_{kin}$ = 100 eV and $\alpha_0$ = 27° (diameter 4.5 Å$^{-1}$). The equipotential contours of the retarding region are illustrated in red. The field of $F$ = -38 V/mm causes slow electrons with energies up to 3.8 eV to undergo a turn within the first 100 μm from the surface. This effectively terminates the Coulomb repulsion between the slow electrons and the photoelectrons of interest. Given the proximity of the conductive surface, the image charge serves to partially shield the Coulomb forces. Accordingly, this regime is designated the dipole regime. For further details, please refer to Refs. [10,32].

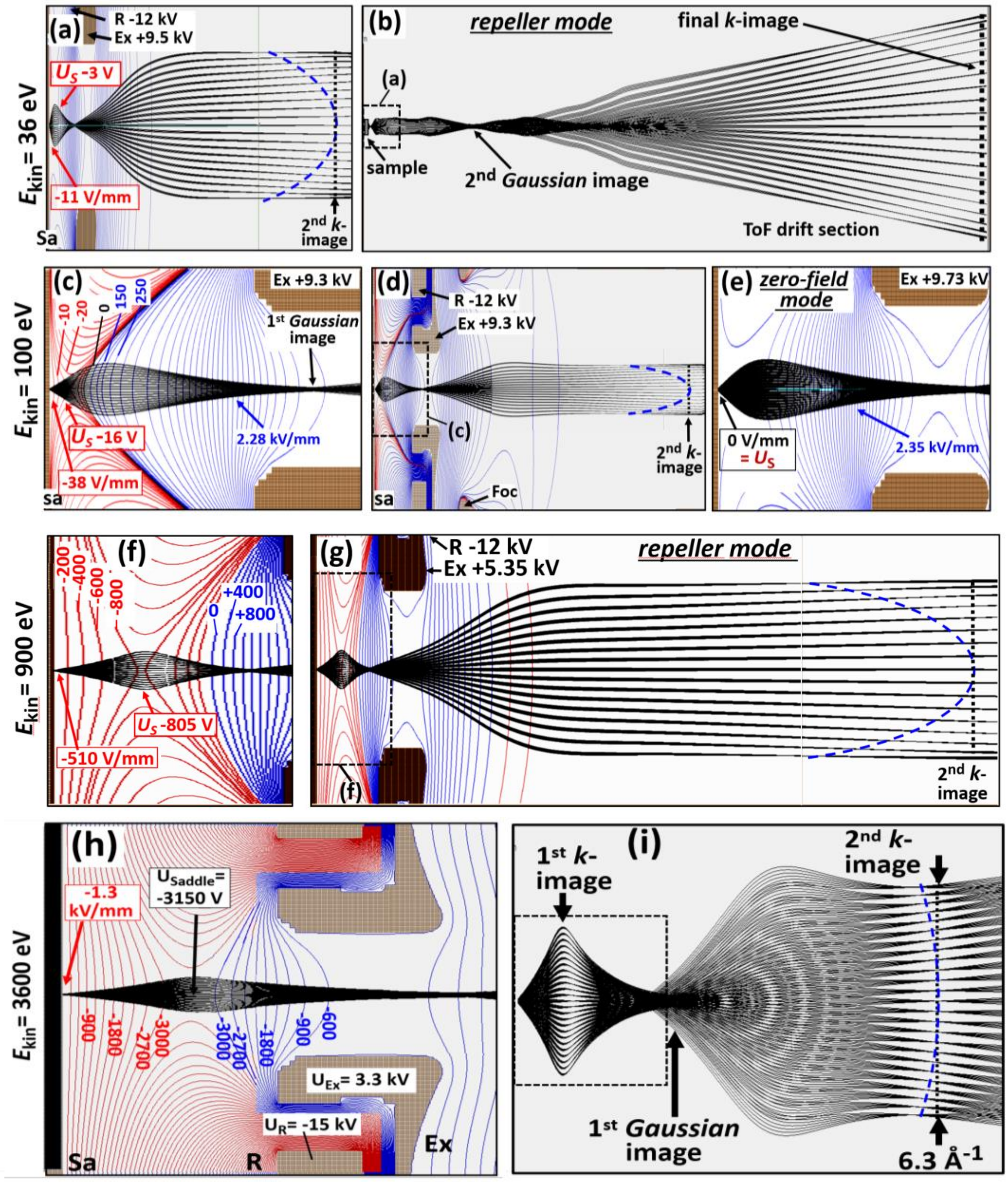

**Fig. 3.** Ray tracing for the *repeller mode* and *zero-field mode*; retarding field $F$ at the sample surface and potential of the saddle point $U_S$ are stated in the panels. (a,b) *Repeller mode* for $E_{kin}$ = 36 eV and angular range 0°-30°, corresponding to a *k*-field of 3 Å$^{-1}$ diameter. (a) Detail from sample (Sa) to the 2$^{nd}$ *k*-image; (b) Simulation for the full ToF MM optics. (c,d) Same for $E_{kin}$ = 100 eV for 0°-27° (*k*-field 4.5 Å$^{-1}$). Retarding and accelerating equipotential contours are marked by red and blue lines, respectively; the numbers represent the potential in volts. (e) *Zero-field mode* obtained by increasing $U_{Ex}$ by $U_{Ex}$ by 430 V. (f,g) *Repeller mode* for $E_{kin}$ = 900 eV and 0°-8° (*k*-field 4.2 Å$^{-1}$). (h,i) Same for $E_{kin}$= 3.6 keV and 0°-6° (diameter 6.3 Å$^{-1}$).

An increase in $U_{Ex}$ to +9.73 kV results in a shift of the saddle point to the sample surface, thereby establishing the *zero-field mode* (see Fig. 3(e)). In these conditions, the field increases gradually and remains relatively low for the first half mm. This is beneficial for samples comprising small three-dimensional structures, such as operando devices with electrodes on the upper surface.

With increasing kinetic energy, the basic behavior remains the same, as can be seen in the results for 900 and 3600 eV shown in Figs. 3(f,g) and (h,i), respectively. To maintain the high refractive power at higher kinetic energies, $U_{Ex}$ is reduced and $U_R$ is increased (for $E_{kin}$= 3.6 keV, $U_R$= -15 kV and $U_{Ex}$= +3.3 kV). We observe a trend that the saddle point and the first *k*-image move closer to the extractor hole and the usable *k*-field diameter increases. The curvature of the equipotential contours and the strong deceleration, e.g. in Fig. 3(h) from 3.6 keV at the sample to 450 eV at the saddle point, act as a strong converging lens. The distance between the object (sample) and the first Gaussian image is ranging between 5 and 10 mm.

The lens geometry is designed so that the position of the second *k*-image coincides with that of the first *k*-image in the gap lens and extractor modes. This plane contains the array of contrast apertures and the alignment grid. As the kinetic energy increases, the curvature of the second *k*-images decreases, as illustrated by the dashed blue curves in Figs. 3 (a,d,g,i). In the same sequence, the usable *k*-field increases from 3 to 6.3 $Å^{-1}$.

Prior to the development of the multi-mode lens, we gained initial experience utilizing the so-called long-range modes, which are a feature of any MM or PEEM with an electrostatic cathode lens. When the extractor is at the same potential as the sample, the *long-range zero-field mode* is formed. This mode has been exploited when tilting the sample in order to record patterns at large off-normal angles (for further details, please refer to Fig. 2(c) in [17] and Figs. 5(j-p) in [19]). When the extractor is set at a negative potential with respect to the sample, the *long-range repeller mode* is yielded. The functioning of this mode was previously corroborated in an experiment conducted at the free-electron laser FLASH at DESY [32], and it has been successfully employed for probing sub-picosecond magnetization dynamics [33].

We summarize the results of Fig. 3 by recalling that the *repeller mode* has been developed for reducing the space-charge interaction. In a multitude of experiments the primary source of Coulomb repulsion exerted on photoelectrons is attributed to slow secondaries and pump-released electrons, particularly in the presence of intense emission from hot spots. The number of slow electrons can exceed that of photoelectrons by orders of magnitude. In order to achieve an optimised lens design, the criterion was established that the *repeller mode, zero-field mode, gap-lens mode* or *extractor mode* can be selected simply by applying different potentials to the front-lens elements. The lens geometry and the position of the contrast aperture mechanism can be kept fixed. The initial experimental outcomes at a HHG source [13] provide evidence that the repeller mode remains effective even at relatively low energies, such as $E_{kin}$ = 16 eV. However, the recorded *k*-field diameter is reduced and the chromatic aberration is enhanced. Therefore, the repeller mode will be employed primarily in instances where space charge effects constitute a significant impediment.

### *2.4 Real-space imaging and ToF X-PEEM mode*

The ability to record high-quality PEEM images using momentum microscopes is important for a number of reasons. It is evident that PEEM represents an effective means of optimising the photon footprint on the sample and the spatial overlap of the pump and probe beams; an example is shown in Ref. [11]. Secondly, the identification of homogeneous regions for *k*-imaging is of particular significance in the context of cleaved samples. Thirdly, the detection and subsequent elimination of emission hot spots from the region of interest (ROI) through a lateral shift is of paramount importance in pump-probe experiments. The intense electron emission from hot spots represents a major source of space charge effects. Fourthly, PEEM imaging is crucial for a *small-ROI mode* of momentum microscopy, analogous to *sub-micron ARPES*. The region of interest (ROI) can be confined by means of a small field aperture in the intermediate Gaussian image. The effective emission region when capturing the *k*-image is restricted by this aperture, irrespective of the photon footprint. ROIs as small as 700 nm in diameter have been selected for *k*-imaging using UV-laser excitation, with a photon footprint of approximately 20 µm [34]. Even smaller ROIs seem feasible, provided that the photon flux in the ROI is sufficiently high.

The aforementioned four applications are supporting measurements of *k*-distributions. Furthermore, the ability to perform high-quality real-space imaging facilitates efficient analysis of the surface chemistry. In a pioneering paper, Tonner [35] proposed a hemisphere-based setup for energy-filtered PEEM in the X-ray range (XPEEM). This concept was subsequently demonstrated experimentally by Bauer and coworkers [36] using synchrotron radiation. Later, the method was extended to higher energies using a double-hemisphere setup designated as HAXPEEM [9]. These approaches are two-dimensional recording methods. A ToF instrument enables the recording of three-dimensional data stacks (x,y,$E_{\mathrm{kin}}$), thereby facilitating the rapid mapping of the chemical structure of the sample surface. The high chemical sensitivity of ToF-XPEEM allows for the determination of not only the atomic species present in the probe volume, but also their chemical states via chemical shifts. In the present study, we examined the real-space imaging performance of the gap-lens and repeller modes for energies within the EUV, soft and hard X-ray ranges. It is of particular interest to examine simulations that elucidate XPEEM with small regions of interest.

Figures 4(a-c) illustrate the results of a PEEM simulation with $E_{\mathrm{kin}}$ = 100 eV in *gap lens mode.* The goal of this simulation was to achieve the maximum possible field of view, which was found to be greater than 4 mm. This diameter is five times larger than that achievable with a standard low-energy momentum microscope. Such large fields are of paramount importance for the forthcoming novel PAXRIXS methodology, wherein the inelastically scattered photons are transformed into photoelectrons by a thin foil [12]. As a consequence of the strong refractive power of the gap lens the ray bundles are contracted and the outer bundles are drawn closer to the optical axis, Fig. 4(a). The field of view is, in fact, slightly larger than the bore of the extractor. A detailed examination of the first Gaussian image in Fig. 4(b) reveals that the central region is well focused, while the peripheral rays appear to be overfocussed. The diameter for which the configuration yields a usable image is approximately 4.2 mm.

Figure 4(c) illustrates the trajectories for the complete microscope optics configuration. At these settings, the lateral magnifications in the first and second Gaussian images are as low as M = 1.1 and 2.6, respectively. This allows for a small final magnification of $M_{tot}$ = 17, enabling the observation of a 4.7 mm FoV on a detector with an 80 mm diameter. In addition to the specific PAXRIXS application, a large FoV is also advantageous for rapid large-area surface inspection or the identification of the photon footprint, for instance following the realignment of the beamline or the setup.

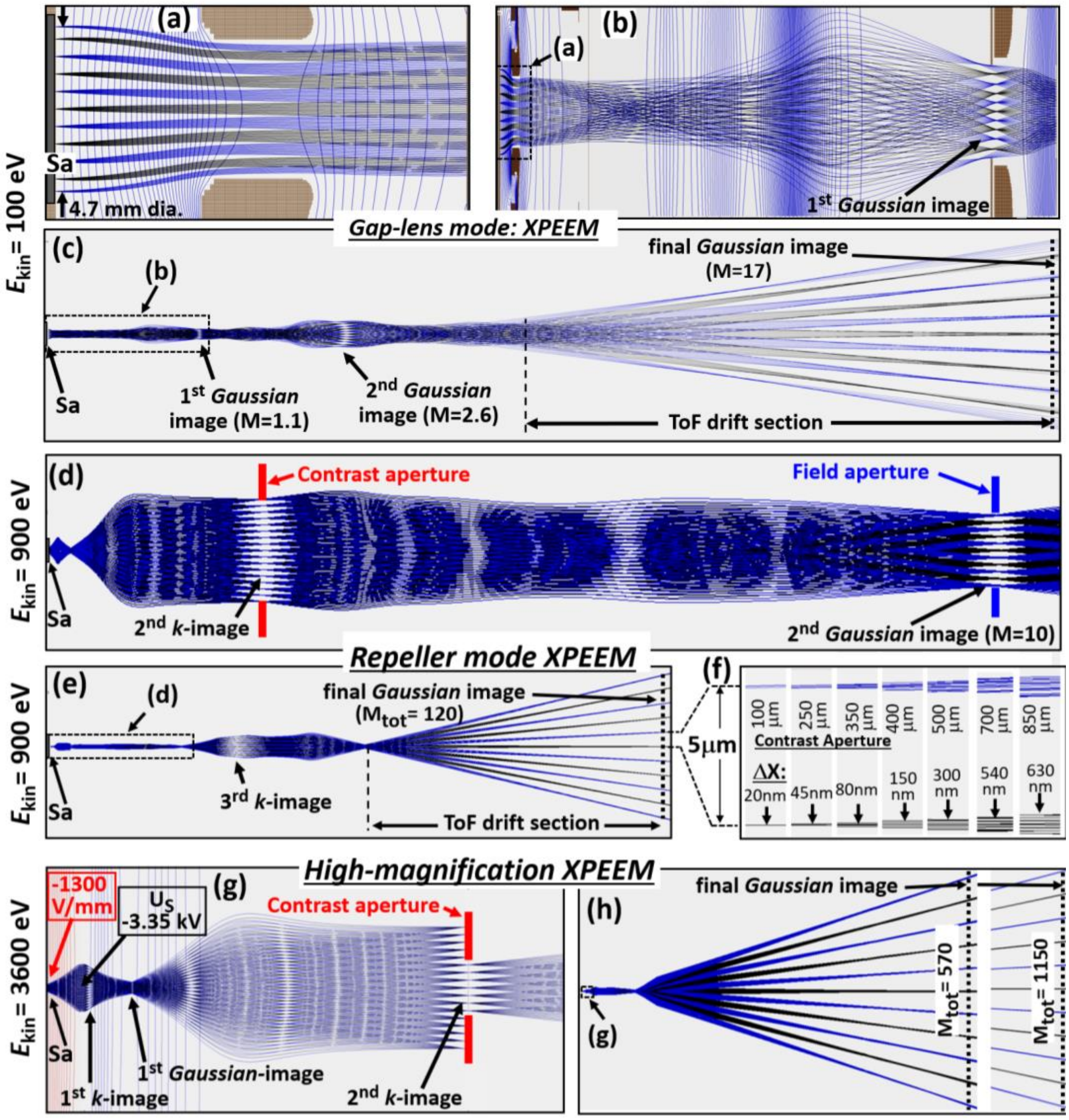

**Fig. 4.** Time-of-flight XPEEM simulations for $E_{kin}$ = 100 eV, 900 eV and 3.6 keV. (a-c) Ray tracing for 100 eV and maximum field-of-view using the *gap-lens mode*. (a) Front section of the lens and (b) first zoom lens up to the 1st Gaussian image. (c) Complete microscope optics showing the 1st and 2nd intermediate images at low magnification (M= 1.1 and 2.6) and the final image at $M_{tot}$= 17. (d-f) Simulation for 900 eV utilizing the *repeller mode*. (d) Upstream section up to the 2nd Gaussian image and (e) complete microscope optics up to the final Gaussian image. (f) Profiles of the two inner ray bundles (distance 5 μm at the sample) for varying diameters of the contrast aperture, with ΔX denoting the diameter of the ray bundle. (g,h) Simulation for $E_{kin}$ = 3.6 keV with strong retarding field of -1.3 kV/mm at the surface and a saddle point of -3.35 kV.

The spatial resolution of XPEEM at synchrotron sources is ultimately limited by space charge effects, as initially identified by Locatelli et al. [7]. This also happens with aberration-corrected instruments [8]. The secondary background rises when $E_{kin}$ is increased. A ToF-MM experiment at $E_{kin}$ = 1 keV using synchrotron radiation from a soft X-ray beamline provided a quantitative assessment of this phenomenon [10]. A shift of 10 eV or more indicates the presence of a macrocharge comprising $10^5$-$10^6$ electrons trailing behind each photoelectron. The Coulomb repulsion is sustained over an extended time period and acts along a macroscopic distance. In the case of metallic samples, the secondary electron spectrum resulting from keV primary electrons exhibits a peak at approximately 2 eV and a width of several eV, which is dependent on the work function (for further details, refer to Ref. [37]). Furthermore, core-level signals with high photoemission cross sections can contribute to the total photoelectron yield. The elimination of slow electrons in the near-surface region should render the *repeller mode* particularly advantageous for XPEEM applications.

Figures 4(d-f) depict the results of a simulation for $E_{kin}$ = 900 eV, wherein Ex and R were set to +5.35 and -12 kV, respectively. These settings are identical to those used in Figs. 3(f,g) and yield a retarding field of -510 V/mm and a saddle point of -805 V. Prior to this experiment, it was unclear how the special potential distribution of the repeller mode would affect the Gaussian image. The trajectories in XPEEM mode up to the second and fourth (= final) Gaussian image are illustrated in Figs. 4(d) and (e), respectively. The positions of the contrast aperture and field aperture, which serve to confine the *k*-range and real-space range, respectively, are indicated in (d). Utilizing a field aperture with a diameter of 500 $\mu$m, the second Gaussian image (with M=10) exhibits a FoV at the sample of 50 $\mu$m. The magnification in the final image is set to $M_{tot}$ = 120.

Figure 4(f) provides a detailed representation of the two central ray bundles, which start at a distance of 5 $\mu$m at the sample. Upon reducing the diameter of the contrast aperture CA from 850 $\mu$m to 100 $\mu$m the cross-section of the central ray bundle exhibits a notable reduction, shrinking from $\Delta$X = 630 nm to 20 nm. The relationship between $\Delta$X and CA is non-linear, due to the fact that spherical aberration is proportional to the third power of the emission angle $\alpha$. The simulations indicate that an XPEEM resolution of approximately 20 nm can be attained even in *repeller mode*. The prerequisite is a sufficiently high photon flux density in a small spot. A spot size of approximately 6 $\mu$m has been achieved at the hard X-ray beamline P22 of PETRA III (Hamburg) using a focusing capillary. It should be noted that the Runge-Kutta algorithm in SIMION [25] produces only approximate results. In a preliminary ToF XPEEM pilot experiment conducted in gap-lens mode, we have achieved resolutions the range of 150 nm with a setup that was not optimized for PEEM [38,13].

Figures 4(g) and (h) illustrate the XPEEM mode for $E_{kin}$ = 3.6 keV, with $U_{Ex}$= 2.6 kV and $U_R$ = -14 kV; similar parameters as utilized in Figs. 3(h,i). The repeller field of -1.3 kV/mm and saddle point at -3.35 kV cut off all electrons except for the topmost 250 eV below $E_F$. This encompasses the majority, if not all, of the core level signals, along with their associated Tougaard background [39,40]. The first *k*-image is close to the saddle point, the first Gaussian image in the centre of the extractor hole and the second *k*-image in the plane of the contrast aperture array. Here we explored the conditions for maximum lateral resolution, which demands a large magnification. Therefore, we assumed a ToF drift tube twice as long as in the previous cases. At a given magnification the spatial resolution of the delay-line detector (approximately 40 $\mu$m) sets a limit for the achievable resolution. For the two simulated cases in Fig. 4(h), $M_{tot}$ = 570 and 1150, we derive resolution values of 70 nm and 35 nm and corresponding fields of view of approximately 20 and 10 $\mu$m, respectively. $k_{\parallel}$ was limited by contrast apertures of 500 and 350 $\mu$m for the M= 570 and 1150 simulations, respectively, as sketched in Fig. 4(g).

XPEEM experiments at keV energies are of interest due to the large mean free path of the photoelectrons, which allows for the study of buried structures and thin-film devices even in operando [9,41]. In such experiments, the *repeller mode* with a strong retarding field, as illustrated in Figs. 4(g) and (h), will be advantageous.

## 3. Summary and Conclusion

A novel front lens has been developed for photoelectron momentum microscopes and PEEMs. The primary goal was to reduce the field strength at the sample surface, thus mitigating the risk of field emission or flashovers. These can occur due to local field enhancement at sharp corners or protrusions of cleaved samples. A second objective was to modify the field distribution in a manner that would result in a retarding field at the surface, thereby redirecting slow electrons back towards the surface. The removal of all slow electrons within a distance of typically less than 100 µm from the surface effectively terminates the major contribution to space charge effects.

Systematic ray tracing calculations were conducted for various geometries, resulting in a lens design that allows for four distinct operational modes through the variation of lens potentials. The *extractor mode*, which operates with a high accelerating field, the *gap-lens mode*, which operates with a significantly reduced field, the *zero-field mode*, which tunes the field at the sample to zero, and the *repeller mode*, which operates with a retarding field at the surface. The requisite additional degree of freedom for field shaping is introduced through the incorporation of one or two annular electrodes, which are concentric and coplanar with the extractor. By varying the voltages of these electrodes, the potential distribution in the gap between the sample and the electrodes can be modified, and the field at the sample surface can be tuned over a wide range, from strongly accelerating to strongly retarding.

When all three electrodes are set at a high positive potential, a homogeneous extractor field is produced, which is the common characteristic of cathode lenses. This mode is well-understood, and its properties have been extensively analyzed [1-5], including the development of analytical expressions for aberrations [6]. By setting the annular electrodes to a smaller potential, it is possible to reduce the field at the sample from its typical value of $F$ = 3 to 10 kV/mm for the extractor mode to values between 0.4 and 1 kV/mm. Consequently, the probability of field emission or flashovers is markedly diminished, as these phenomena frequently exhibit threshold behavior as a function of $F$. The configuration of the extractor electrode in the center at a high positive voltage, surrounded by the annular electrodes at smaller voltages, forms a *gap lens* between the sample and the electrodes. Despite the markedly diminished field at the sample, the aberrations of the gap lens are small. The field curvature of the $k$-image in the backfocal plane is significantly reduced, and the depth of focus is larger than that of a conventional cathode lens in extractor mode. These advantages are even more significant at higher kinetic energies.

The proximity of the principal plane of the gap lens to the sample is beneficial for achieving a high collection efficiency. This enables the imaging of $k$-fields as large as 18 Å$^{-1}$, which is crucial for XPD experiments with soft or hard X-ray excitation [18,19]. The ability to image fields of view up to 4 mm in diameter in XPEEM mode will be advantageous for the upcoming technique of PAXRIXS [12]. The gap lens's high refractive power and proximity to the object evoke the characteristics of an objective lens in an optical microscope.

Subsequently, an aperture lens (the extractor hole) follows the gap lens, exhibiting negative aberration coefficients [42]. In contrast to optical microscopy, the use of round lenses in particle optics precludes the possibility of eliminating spherical aberration, as demonstrated by Scherzer and Rose [43,44].

The application of a negative potential to the annular electrode(s) enables the adjustment of the field at the sample to zero and the minimization of the field gradient, which is advantageous for non-planar samples. The application of a retarding potential at the sample surface deflects all slow electrons back to the surface, thereby suppressing the primary cause of space charge effects. With increasing photoelectron kinetic energy, the number of slow secondary electrons increases markedly. In pump-probe experiments, a second species of slow electrons is generated via nPPE processes by the pump pulse. The elimination of the primary contributor to Coulomb forces thus paves the way for a promising future for the *repeller mode* in time-resolved pump-probe experiments. In light of the involvement of both high photoelectron energies and pump-released slow electrons, this is of particular importance for experiments at X-ray free-electron lasers.

For the same reason of space charge reduction, this mode is advantageous for real space imaging at high energies (XPEEM), which is limited by space charge effects [7-9]. Systematic ray-tracing calculations provide a quantitative basis for previous general considerations regarding the reduction of space charge effects, as discussed in Refs. [10,26,32]. The benefits of the repeller mode are accompanied by an increase in chromatic aberration. This is a pertinent consideration in the context of ToF-MM in the valence range, given that an energy interval of several eV is recorded simultaneously. The lateral chromatic aberration, or the "breathing" of the image as it varies in energy, can be corrected numerically [23]. However, the longitudinal chromatic aberration (shift of the focal plane) cannot be corrected and necessitates the acquisition of multiple (smaller) energy intervals in a sequential manner. This is not a significant issue for core level XPD and XPEEM, as the width of the signals is typically less than 1 eV.

In conclusion, the novel front lens configuration offers a variety of operational modes that are not possible with conventional cathode lenses. In particular, two undesirable phenomena make these modes attractive: A high extractor field may cause field emission or flashovers, which may impose constraints on the maximum usable field. Furthermore, the presence of a large number of slow electrons may result in the emergence of space-charge effects, which can potentially impair the energy and *k*-resolution. To date, the novel modes have been employed in a limited number of experiments, which are summarized in Ref. [13]. In experiments conducted with a HHG-based source at a low kinetic energy of 16 eV, the gap lens and repeller modes have been demonstrated. The gap lens mode has been shown to perform well when used with soft X-rays at the synchrotron source PETRA III [27,28]. An initial iteration of the repeller mode (in which the extractor served as the repeller electrode) was investigated at the free electron laser FLASH at DESY [32] and subsequently employed in experiments [33].

## Acknowledgements

We thank H.-J. Elmers, O. Fedchenko and Y. Lytvynenko (Mainz Univ.), S. Beaulieu, S. Fragkos and Y. Mairesse (CELIA, Bordeaux), T. Allison and A. Kunin (Stony Brook and Princeton Univ.), S. Chernov, C. Schlueter, M. Hoesch, D. Kutnyakhov and M. Scholz (DESY, Hamburg), Q. Nguyen (SLAC, Stanford) and B. Schönhense (Springbok.ai) for fruitful discussions on various aspects of this article. This development was funded by the BMBF (Projects 05K22UM1, 05K22UM2 and 05K22UM4) and Deutsche Forschungsgemeinschaft DFG (German Research Foundation) through TRR 173-268565370 Spin +X (project A02) and Project No. Scho341/16-1.

## Data availability

All data shown within this article is available on reasonable request. We declare no conflict of interest.